\documentstyle[preprint,aps,epsf]{revtex}
\tighten
\begin{document}
\draft
\title{Cauchy Horizon Stability and Cosmic Censorship}
\author{Wenceslao S. German and Ian G. Moss}
\address{
Department of Physics, University of Newcastle Upon Tyne, NE1 7RU U.K.
}
\date{March 2001}
\maketitle
\begin{abstract}
Some interesting consequences of spacelike matter shells are presented, in
particular the possibility of travelling through Cauchy horizons and violating
the strong cosmic censorship hypothesis. These show that the weak energy
condition does not guarentee cosmic censorship.
\end{abstract}  
\pacs{Pacs numbers:  04.20.Dw 04.70.Bw}  
\narrowtext

The ability to predict the future from given initial conditions might seem 
like an essential requirement for a classical theory of physics, but this
remains an open issue in General Relativity whilst the conditions needed for
cosmic censorship are not known. The situation could be resolved by finding
some property of realistic matter which prevents the formation of naked
singularities. In this letter we examine this question by looking at
violations of cosmic censorship inside black holes.

The form of the strong cosmic censorship conjecture which we use is that
`every generic, inextendible space-time containing physically reasonable
matter is globally hyperbolic' \cite{clarke}. This form of the conjecture
seems to be difficult to violate. If we take the charged black hole
solution, for example, this is extendable beyond the globally hyperbolic
region by the usual coordinate construction, but it is not considered generic
because the Cauchy horizon is not stable to linear perturbations.

The instability of the Cauchy horizon has a simple physical explanation
originally due to Penrose \cite{penrose}. Incoming radiation ariving at the
Cauchy horizon is blue shifted and the energy flux measured by an observer
approaching the Cauchy horizon diverges \cite{chandra,chandra1}.

A detailed analysis of black hole perturbations in spacetimes with a
cosmological constant has shown that the stability of the Cauchy horizon can
still be related to the energy flux. Stability depends on the values of the
surface gravity at the Cauchy horizon $\kappa_1$, the event horizon $\kappa_2$
and the cosmological horizon $\kappa_3$. Stability requires
$\kappa_1<\kappa_3$ and $\kappa_1<\kappa_2$.
\cite{mellor,brady,chambers,brady1,brady2}. 
(The second requirement was not
appreciated before reference \cite{brady2}).

None of the vacuum black hole spacetimes satisfy the stability requirements.
We will consider the effect of adding matter, specifically spacelike shells,
to the inside of the black hole. These spacelike shells are best thought of
as transition layers separating different vacuum phases, where the phase
transition is triggered by the high spacetime curvature inside the black hole
\cite{shore}. The shells can have internal stresses and carry currents, and
generalise the bubble walls associated with broken symmetry phase transitions
\cite{blau}.

Similar forms of matter have appeared in the literature previously in
connection with the limiting curvature hypothesis, the idea that quantum
gravity effects may prevent spacetime curvature singularities \cite{frolov}.
The spacelike shells allow the transformation of the black hole into a
nonsingular wormhole. However, we shall consider only black holes with
singularities.

\begin{figure}
\begin{center}
\leavevmode
\epsfysize=20pc
\epsffile{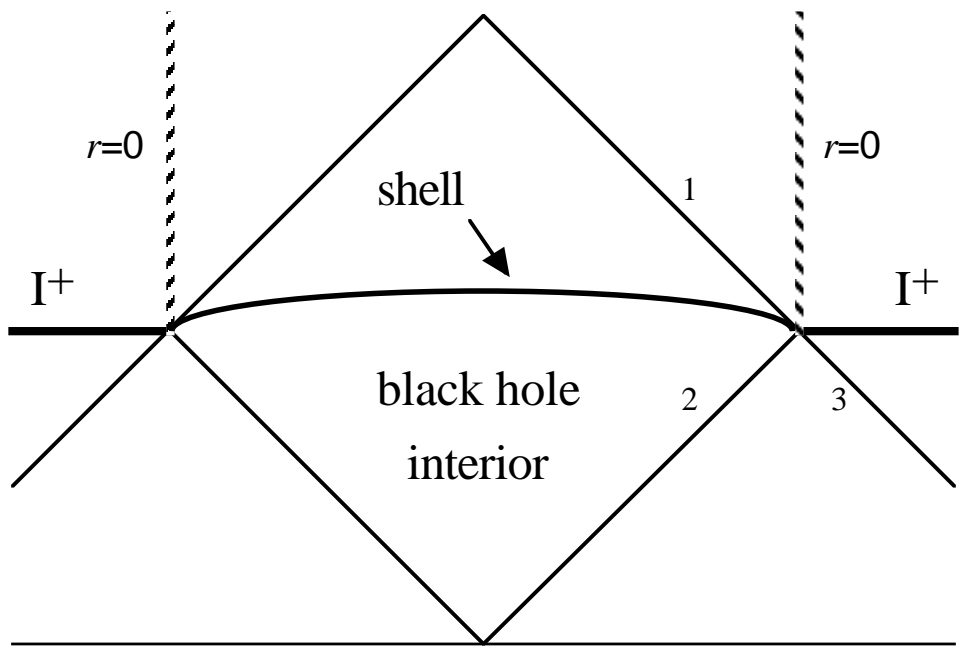}
\end{center}
\caption{Penrose diagram of the charged black hole with inner horizon $r=r_1$,
event horizon $r=r_2$ and cosmological horizon $r=r_3$. The shell lies between
the inner and the event horizons.} 
\label{fig1}  
\end{figure}

The spacetime is shown in figure (\ref{fig1}). The shell is placed at $r_s$
and the metric on either side of the shell is given by
\begin{equation}
ds^2=-{r^2\over\Delta}dr^2+{\Delta\over r^2}dt^2+r^2\Omega^2\label{metric}
\end{equation}
where
\begin{equation}
\Delta=2Mr-r^2+Q^2-\case1/3\Lambda r^4\label{delta}
\end{equation}
For $r<r_s$, the parameters take on values $M_1$, $Q_1$ and $\Lambda_1$ and for
$r>r_s$, $M_2$, $Q_2$ and $\Lambda_2$. Both $\Lambda_1$ and $\Lambda_2$ will be
suposed fixed by the details of some phase transition. For simplicity, we will
take $Q_2=0$. This would be the case, for example, if the charge $Q_1$ is
associated with a gauge symmetry which is broken outside the shell.

\begin{figure}
\begin{center}
\leavevmode
\epsfxsize=30pc
\epsffile{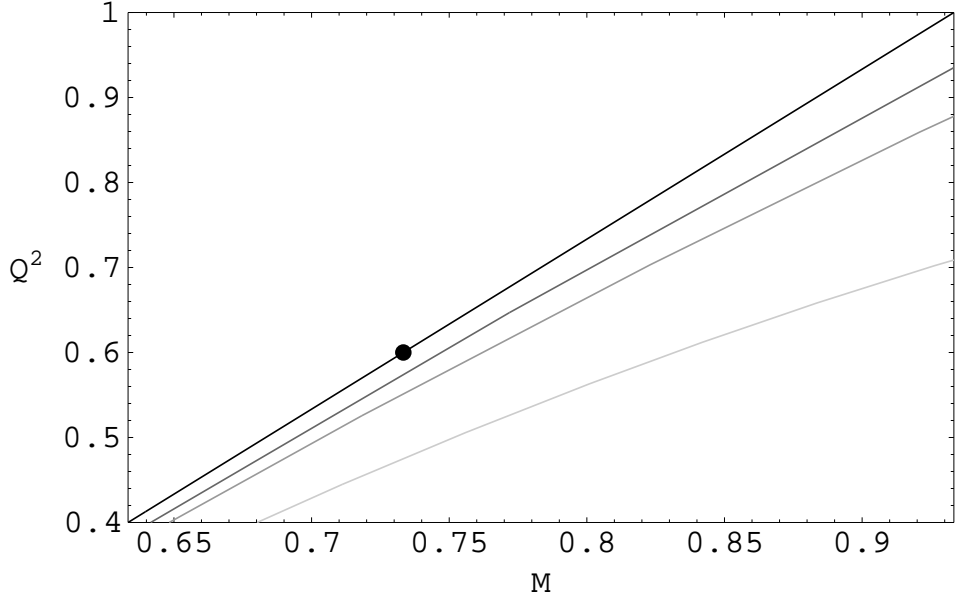}
\end{center}
\caption{Contours of constant shell pressure $p_\perp$ have been
drawn on the charge and mass paramater space of the inner black
hole solution. Lighter gray indicates smaller pressure. The
straight line is the limiting case $p_\perp\to\infty$. (the
parameters are scaled by the horizon radius, $M=M_1/r_2$,
$Q=Q_1/r_2$ and $\Lambda_1r_2^2=0.4$)}
\label{fig2} 
\end{figure}

Suppose that $h_{ab}$ is the metric induced in the shell and $u_a$ is a unit
vector along the $t$ direction, then the stress-energy of the shell has the
form \cite{frolov,balbinot}
\begin{equation}
S_{ab}=p_s u_a u_b+p_\perp(h_{ab}-u_a u_b).
\end{equation}
The junction conditions imply that the metric is continuous and the extrinsic
curvature satisfies,
\begin{equation}
\left[K_{ab}\right]=-8\pi\left(S_{ab}-\case1/2 h_{ab}S\right).
\end{equation}
For the metric (\ref{metric}),
\begin{eqnarray}
\left[\Delta^{1/2}\right]&=&4\pi r^2 p_s,\\
\left[r(\Delta^{1/2})'\right]&=&8\pi r^2p_\perp.
\end{eqnarray}
The junction conditions can be solved for $M_1$ and $Q_1$ as functions of the
radius $r_2$ and the pressures $p_s$ and $p_\perp$. Figure (\ref{fig2}) shows
contours of constant pressure $p_\perp$ when $p_s=0$. As $p_\perp\to\infty$,
the solution approaches a line
\begin{equation}
Q^2-Q_c^2=2r_2\left(M-M_c\right)
\end{equation}
where $M_c=r_2-\case2/3\Lambda_1r_2^2$ and $Q_c^2=r_2^2-\Lambda r_2^4$
correspond to a black hole with coincident horizons.

The linear perturbation analysis of the black hole can be carried out on
either side of the shell. The perturbations propagate as waves and undergo a
constant redshift at the shell, leaving the same Cauchy horizon stabilty
conditions on the surface gravity as before.

\begin{figure}
\begin{center}
\leavevmode
\epsfxsize=30pc
\epsffile{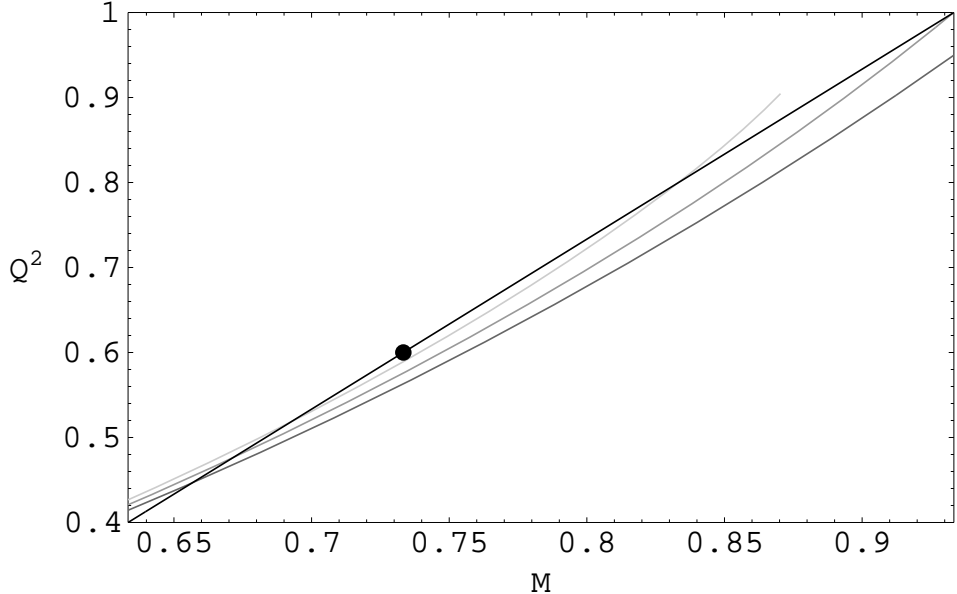}
\end{center}
\caption{Contours of constant surface gravity $\kappa_1$ have been drawn on the
charge and mass parameter space of the inner black hole solution. The straight
line is the limiting case of the shell pressure $p_\perp\to\infty$.}
\label{fig3}  \end{figure}

The Cauchy horizon lies inside the shell at $r=r_1$, where the surface gravity
is given by $\kappa_1=\Delta'/(2r^2)$. From (\ref{delta}), and the condition
$\Delta(r_1)=0$, one can obtain
\begin{eqnarray}
M_1&=&\kappa_1r_1^2+r_1-\case2/3\Lambda_1r_1^3\\
Q_1^2&=&2\kappa_1r_1^3+r_1^2-\Lambda_1r_1^4
\end{eqnarray}
Figure (\ref{fig3}) shows contours of constant $\kappa_1$ in the $(M_1,Q_1)$
parameter space. The surface gravity vanishes at the point $(M_c,Q_c)$.
Comparing this figure with figure (\ref{fig2}), we see that if the
pressure is sufficiently large, we can make the surface gravity arbitrarily
small. In particular, we can satisfy the stability requirements
$\kappa_1<\kappa_2$ and $\kappa_1<\kappa_3$ for the Cauchy horizon. 

The stability of the shell itself can be analysed by taking the radius $r_s$
to be a function of the proper time along the shell, as described in reference
\cite{balbinot}. The junction conditions become
\begin{eqnarray}
\left[K_{\theta\theta}\right]&=&4\pi r^2p_s\\
\left[\dot K_{\theta\theta}\right]&=&8\pi r\dot r p_\perp
\end{eqnarray}
where $K_{\theta\theta}=(r^2\dot r^2+\Delta)^{1/2}$. If $\delta p_s=c^2\,\delta
p_\perp$, perturbations about the static shell satisfy
\begin{equation}
\left[\Delta^{-1/2}\right]\delta \ddot r+
\left({1\over 2}\left[r^{-1}(\Delta^{-1/2}r^{-1}\Delta')'\right]
-{2\over c^2}\left[r^{-1}(r^{-2}\Delta^{1/2})'\right]\right)
\delta r=0
\end{equation}
We find neutral stability when $p_s=0$, whilst for $0\le p_s\ll p_\perp$ there
exists a $c_{max}$ such that stability occurs for $0<c<c_{max}$.

The spacelike shells which we have considered have zero density and therefore
when $p_s\ge 0$ they marginally satisfy the weak energy condition, $\rho\ge0$
and $\rho+p\ge 0$. We can infer that the weak energy condition does not imply
strong cosmic censorship.

The spacelike shells considered by Frolov et al. \cite{frolov} were
specifically aimed at domonstrating the possibility of singularity avoidance
in gravitational collapse by the creation of wormholes and for this the shell
must have `exotic' matter with $p+\rho<0$. Wormholes are not globally
hyperbolic and violate the strong cosmic censoship principle. We find that
spacelike shells which satisfy the weak energy condition can still violate
cosmic censorship and produce naked singularities.  

\acknowledgments

We would like to thank Manolo Per for helpful discussions. WSG is supported by
CONACYT (Mexico) grant number 116020.

\end{document}